\documentclass[
]{ceurart}

\usepackage{subfig}
\usepackage{algorithm}
\usepackage{multirow}
\usepackage{caption}
\begin{document}

\copyrightyear{2021}
\copyrightclause{Copyright for this paper by its authors.
  Use permitted under Creative Commons License Attribution 4.0
  International (CC BY 4.0).}

\conference{Perspectives on the Evaluation of Recommender Systems Workshop (PERSPECTIVES 2021), September 25th, 2021,
co-located with the 15th ACM Conference on Recommender Systems, Amsterdam, The Netherlands}

    
\title{Modeling Online Behavior in Recommender Systems: The Importance of Temporal Context}

\author[1,2]{Milena Filipovic}[%
email=milena.filipovic1@swisscom.com
]
\address[1]{Swisscom, Switzerland}
\address[2]{Ecole Polytechnique Fédérale de Lausanne, Switzerland}

\author[2,3]{Blagoj Mitrevski}[%
email=blagoj.mitrevski@symphony.is
]
\address[3]{Symphony, North Macedonia}

\author[2]{Diego Antognini}[%
email=diego.antognini@epfl.com
]

\author[1]{Emma Lejal Glaude}[%
email=emma.lejalglaude@swisscom.com
]

\author[2]{Boi Faltings}[%
email=boi.faltings@epfl.com
]

\author[1]{Claudiu Musat}[%
email=claudiu.musat@swisscom.com
]

\begin{abstract}
  Recommender systems research tends to evaluate model performance offline and on randomly sampled targets, yet the same systems are later used to predict user behavior sequentially from a fixed point in time.
    Simulating online recommender system performance is notoriously difficult and the discrepancy between online and offline behaviors is typically not accounted for in offline evaluations.
    This disparity permits weaknesses to go unnoticed until the model is deployed in a production setting.
    In this paper, we first demonstrate how omitting temporal context when evaluating recommender system performance leads to false confidence.
    To overcome this, we postulate that offline evaluation protocols can only model real-life use-cases if they account for temporal context.
    Next, we propose a training procedure to further embed the temporal context in existing models. We use a  multi-objective approach to introduce temporal context into traditionally time-unaware recommender systems and confirm its advantage via the proposed evaluation protocol.
    Finally, we validate that the Pareto Fronts obtained with the added objective dominate those produced by state-of-the-art models that are only optimized for accuracy on three real-world publicly available datasets.
    The results show that including our temporal objective can improve recall@20 by up to 20\%
\end{abstract}


\maketitle

\section{Introduction}

In an increasingly digital world, 
recommender systems are a staple of our daily routines. They influence how we perceive our environment, from media content to human relationships.
Traditional methods of evaluation that entail random sampling over a long period of time are perfect for a system that is designed to remain unchanged for an equally long and predefined period.
However, if the system is to be used in a dynamic setting, e.g. recommending the next song to play in a playlist, the way it is evaluated must reflect that.
Inadequate evaluation techniques can lead to false confidence, which is especially detrimental in commercial settings.

Recommender system evaluation can be done \emph{online} or \emph{offline}. \emph{Online} evaluation implies deployment of the recommender in the real world, often in a commercial setting. While this may be ideal for measuring the real-life impact of a system, it is also costly, both in terms of resources and time, and therefore rarely used in research and benchmarking.
In \emph{offline} evaluation, historical data is utilized. Some portion of the data is selected to train on, while another subset is used for performance testing. Since all data points are available beforehand, the evaluation costs and overall timeline are significantly reduced.    

Irrespective of whether they are evaluated online or offline, many  existing  recommenders  ignore  temporal  information. Most recommender systems fall into one of two main categories: content-based or collaborative filtering \cite{Zhou_2020,he2017neural,liang2018variational,antognini2020interacting}. 
The former relies on recommended items having similar attributes to those that the user has previously interacted with, while the latter methods base their recommendation on items bought by similar users. 
However, 
many models completely ignore temporal information, with the notable exception of time-aware recommender systems\cite{campos2014time}. These systems introduce additional context to interactions: their temporal dimension.

In this work, we first focus on the importance of temporal dynamics in recommender system evaluation. 
We draw attention to the lack of standardization in the evaluations, and the differences between research settings and the systems' ultimate applications.
Then, we highlight two temporal evaluation protocols and show how they attain a closer approximation of the real-life conditions in which recommender systems are deployed.
Second, we present a multi-objective approach\cite{milojkovic2019multi} of incorporating the temporal context to time-unaware recommender systems  without any change in model architecture.
We introduce a naive recency objective as a means to include temporal dynamics in typically time-independent recommender systems. We also provide a measure of recency in the form of a performance~metric.
Through experiments on three real-world publicly available datasets we show that the addition of the naive temporal objective yields improvements not only in recency but also in relevance. 
Finally, we demonstrate that the  Pareto Fronts  obtained  with  the added objective dominate those produced by state-of-the-art models.

To the best of our knowledge, this is the first study quantifying the difference in recommender system performance when evaluated using methods that model real-world environments, as opposed to traditional techniques. We also show that a recommender system can be optimized for both relevance and recency objectives simultaneously.
To summarize, the main contributions of this paper are as follows:
\begin{itemize}
    \item We demonstrate how commonly used evaluation protocols do not provide adequate modeling of real-world deployment settings. To combat this, we show two evaluation techniques to facilitate offline modeling of online production environments that inherently incorporate temporal dynamics;
    \item We introduce a ``naive'' recency function that can be utilized to create a temporal objective. We show that optimizing for both temporal context and relevance \cite{milojkovic2019multi} leads to solutions that dominate those optimized just for relevance in both dimensions.
\end{itemize}

\section{Related Work}
\subsection{Evaluating Recommender Systems}

\subsubsection{Traditional Recommender Systems.} Inputs and outputs share similarities with classification and regression modeling: a class variable is predicted from a set of given features. Given that recommendation tasks can be seen as a generalization of these, some evaluation techniques used for classification are transferable to recommender systems.

In collaborative filtering research, recommenders are generally evaluated either through \emph{strong} or \emph{weak generalization}, as characterized by \cite{marlin2004collaborative}.
In both approaches, models are trained on observed interactions and validated or tested on those that are held-out. However, there exist some key differences.
\emph{Weak generalization} is introduced in \cite{10.5555/2074094.2074100}, where the held-out set is created through random sampling of the available interactions of all users.
\emph{Strong generalization} differs by taking disjoint sets of users for the training, validation, and testing sets. Following this, some interactions are held-out from the validation and test sets and then approximated using the recommender.
Methods that encode user representation cannot apply \emph{strong generalization}, as they cannot generate outputs for previously unseen users. 
An example of the \emph{strong generalization} approach can be seen in \cite{liang2018variational}, whereas \cite{ning2011slim, wu2016collaborative, rendle2012bpr} all use \emph{weak generalization}.


Several of these works emphasize that the application of their recommender system would be in predicting future user actions, yet all validation and testing is done with randomly selected interactions.
This can break the time linearity as the knowledge of future interactions during training can help predict a randomly sampled past interaction.
While much effort is directed towards establishing the importance of proper evaluation design, it is generally focused on implementing relevant metrics to avoid under- or over-estimating real-world performance \cite{aggarwal2016recommender}, and not on the evaluation procedures themselves.

\subsubsection{Temporal Recommender Systems.} They denote time-aware RS (TARS), and incorporate time explicitly or implicitly\cite{campos2014time}. 

Temporal recommender systems can be taken to include sequence-aware recommender systems (SARS), as a special form of time adaptive recommenders that focus on ordering rather than specific time instances\cite{campos2014time}. 
It is however important to note that while they can be evaluated using similar techniques, SARS approach temporal dynamics from a different perspective, therefore the resulting models can differ greatly from typical TARS\cite{quadrana2018sequence}.
\cite{campos2014time} provide an extensive overview of possible evaluation techniques, which served as an inspiration and point of reference for this work.
While traditional evaluation protocols may be used on temporal recommenders, it is more representative to preserve the temporal ordering between interactions since this is something that the recommender aims to learn. By extension, train, validation, and test splits should also be ordered.
\cite{quadrana2018sequence} state that they were unable to find a consensus among evaluation protocols used in recent sequence-aware recommender work, which is mirrored in our findings. 
Yet we did determine that most recent SARS focus only on next item prediction, meaning they output one recommendation. They also typically employ certain target item conditions to decrease computational cost \cite{campos2014time}. 
The target item conditions determine the (sub)set of items for which a recommender should produce predictions and are specific for top-N recommendation evaluation. 
The reduction of the computational costs is generally done through conditions that rank one ground truth item against a set of other items false items.
Examples can be found in \cite{sun2019bert4rec, kang2018self, hidasi2018recurrent}. We return to the problem of subsampling in Section \ref{sec:proposed}.

\subsection{Temporal Context in Recommender Systems}

In this paper, we introduce the concept of recency.
An important note is that there are multiple definitions of recency in recommender systems literature.
In fact, this lack of consensus has persisted for years.
\cite{ding2006recency, vinagre2015collaborative} treat the recency of an item as an attribute that is user-dependent. The value is determined by the last time the user interacted with a given item. \cite{chakraborty2017optimizing, gabriel2019contextual} also claim to incorporate recency into their research: when recommending news articles, they measure recency as the age of the item on the platform. Our analysis will follow the latter definition. This is in line with our desire to explore the effects of a light-weight temporal addition on the performance of traditional RS. Further work to determine the "ideal" definition of recency, while undoubtedly invaluable, is outside the scope of this work. 

\section{Evaluation Protocols}
\label{sec:proposed}
We propose that the temporal dimension should be considered when evaluating the performance of any recommender.
While random sampling may be an appropriate target selection technique for some classification or regression tasks, we argue that this is not the case when it comes to predicting a user's subsequent move.

Unlike the vast majority of evaluation methods applied to traditional recommenders, temporal recommender systems literature does model the passage of time.
However, as stated above, the performance is often computed over a subset of the itemset and the user's true chosen item.
The argument is that subsampling is necessary due to the complexity of the ranking task. While this has some validity, itemsets of around 10,000 datapoints can be ranked highly efficiently, especially when taking into consideration recent advancements in machine learning libraries and GPU programming. Therefore, we do not utilize subsampling in our work.

The adoption of a recommender system in real scenarios has two major phases. The first, called the development phase, is purely offline and theoretical. In this part, three separate sets of data must be created: a training set that the model will use to learn item and user representations, a validation set for hyperparameter tuning, and a test set to evaluate how well the model performs.
The second, called the deployment-ready phase, include interactions with end-users. The maximum amount of data is leveraged to train a model with as much information as possible, evaluate its performance, and then deploy it into production. In this case, only two sets are needed: training and validation sets.

\begin{figure}[t]
\centering
\subfloat[Proportional Temporal Selection.\label{fig:prop}]{ \includegraphics[width=.45\linewidth]{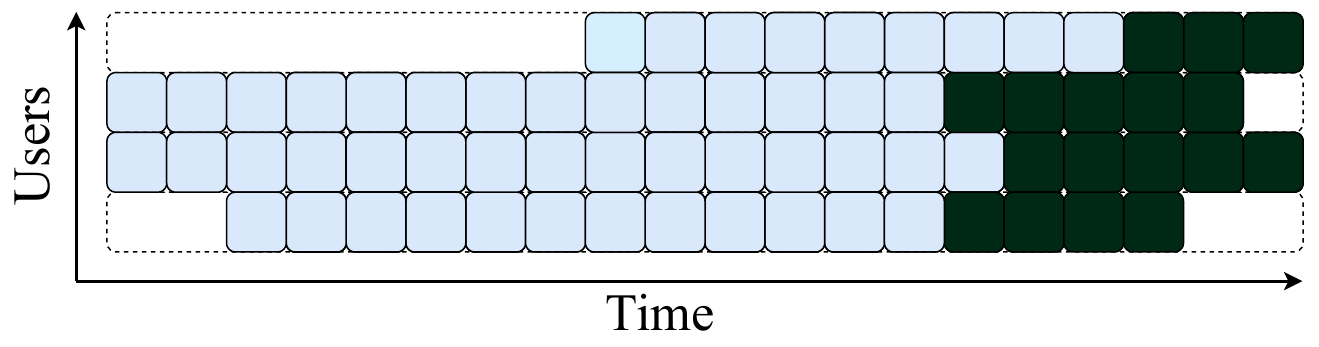}}
\subfloat[Strict Temporal Cutoff. \label{fig:cutoff}]{ \includegraphics[width=.45\linewidth]{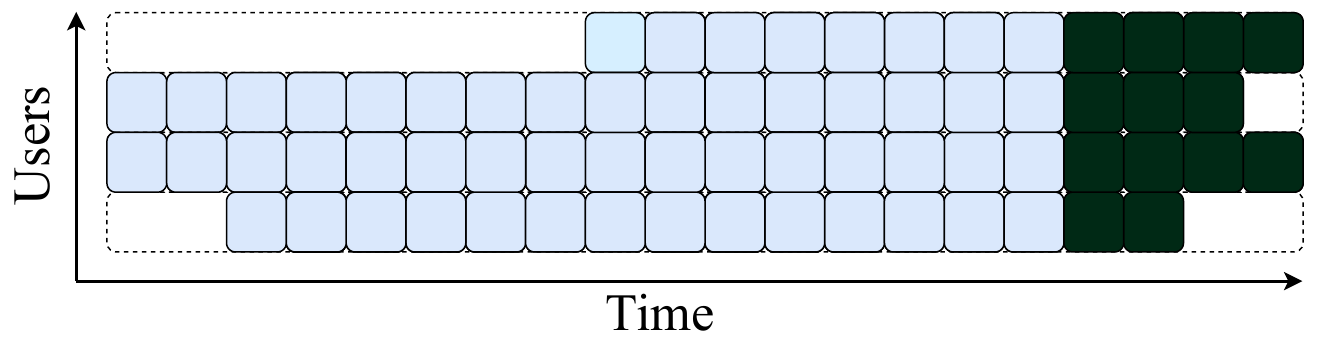}}
\caption{Two methods for temporal validation set target selection.}
\label{fig:TempValFig}
\end{figure}
 
One downside of collaborative filtering methods is that most models are incapable of incorporating new items without retraining. While ways to alleviate this problem have been explored \cite{luo2012incremental}, the issue remains widespread and worthy of more study, but lies outside the scope of this paper.
Therefore, we assume an industry-like environment: the recommender system will be retrained regularly and will be exposed to clients for a relatively short period, ranging from a couple of days to a 
few months. 
We postulate that the performance of the recommender on the last portion of historically available data is most indicative of how it will behave when deployed.

Our protocols focus on set creation. When selecting the target values in a validation set, we take two possible approaches. 
The first, \emph{proportional selection}, depicted in Figure \ref{fig:prop}, selects the final $X\%$ of each user's interactions and uses these to create target items. 
Here we preserve the time ordering of the input and target interactions, maintaining similarity with the real-life use-case. 
The second approach, shown in Figure \ref{fig:cutoff}, is 
based on a \emph{strict time cutoff} to select the target items of the validation set. This method is even closer to the real-world use case. However, it does suffer from certain drawbacks as user interactions are not necessarily evenly distributed through time, leading to some users being more represented than others in the target set.
While these are similar to the suggestions developed in \cite{campos2014time}, we underline that these approaches should not be limited to evaluating TARS.
It is crucial to approximate with maximum precision the performance of a model when developing a novel system, before it is released into production.
The second approach directly models the real-world context and contains user-item interaction sequences created after a specific strict time cutoff.

\section{Recency to Improve Recommendation}
The main task of a recommender system is to anticipate users' future desires and suggest relevant content.
The \emph{relevance} objective is the one that is most commonly found in recommender systems literature and accounts for the accuracy or correctness. It actively focuses the recommender on selecting the item(s) with which the users will most likely interact.

However, just recommending the most relevant items does not always satisfy all the concerns of those building the system and it is not the only objective used in practice. We distinguish two types of objectives: \emph{correlated} and \emph{uncorrelated} to relevance. 
The former ones correspond to those whose optimization is linked to the relevance objective. Examples are novelty \cite{vargas2011rank}, serendipity \cite{ge2010beyond}, and utility-based objectives, such as revenue. 
For the latter, \emph{not correlated to relevance}, examples can be found in diversity and fairness. 
In this work, we introduce a simple utility-based objective used to inject temporal information alongside the relevance objective. While the exploration of uncorrelated objectives is essential for the future of recommender systems, we leave it for future~work.

\subsection{Adding Temporal Context}
Based on our experience with real-life use-cases, we observed that users seem to gravitate towards content that had more recently been added to a given platform. While we cannot disclose internal facts and figures, the temporal objective described below was motivated by behaviour exhibited across many months of user interactions observed internally. 
Building on these findings, and works such as \cite{chakraborty2017optimizing} and \cite{gabriel2019contextual}, we chose to explore the effects of incorporating recency as an objective during the learning phase. Given an item $x$ with a timestamp $t_x$, we further define the recency function $f(\cdot)$ as:
\begin{equation}
f(x) = 
\left\{
\begin{array}{ll}
      1 &  \frac{t_{x} - t_{min}}{t_{max} - t_{min}} \geq 0.8 \\
      0.3^{(0.8 - \frac{t_{x} - t_{min}}{t_{max} - t_{min}})\times \frac{10}{3}}& otherwise \\
\end{array} 
\right.
\label{eq:recency}
\end{equation}
where $t_{max}$ and $t_{min}$ are the maximum (most recent) and minimum (oldest) timestamps over the itemset. In $f(\cdot)$, we first scale all timestamps to $[0, 1]$ using the min-max scaler, and then apply a transformation inspired by \cite{huang2013incorporating}.

We underline that this is a naive function which we found to best approximate user interactions observed in-house. Works such as \cite{kowald2017temporal} rely on a power law distribution to model temporal effects, highlighting that further exploration into use-case specific temporal context approximations may yield exciting results.

The recency objective is formulated as a loss that stimulates the recommendation of recent items.
Each item in the itemset is assigned a recency weight, 
The vector is then used to weigh item importance when calculating the loss. Adding weights into a traditional loss does not affect the differentiability of the function.

To illustrate how our temporal objective can be easily integrated into a traditionally time-unaware RS,
we take as a use-case the state-of-the-art variational autoencoder Mult-VAE\textsuperscript{PR} of \cite{liang2018variational}. For the sake of brevity, we refer the reader to \cite{liang2018variational} for more details about the model.
We thus propose an extension of Mult-VAE\textsuperscript{PR}, where the loss function for user $u$ is modified to: 
\begin{equation}
\mathcal{L}_{\beta}(x_u;\theta, \phi) = \mathbb{E}_{q_\phi(z_{u}|x_{u})}\left[ \log p_\theta(f(x_u) \ast x_{u}|z_{u})\right] \\
- \beta \cdot \mathrm{KL}(q_\phi(z_{u}|x_{u}) || p(z_{u})) 
\end{equation}
where the expected negative log-likelihood is modified to include the element-wise multiplication of input vector $x_u$ by $f(x_u)$, which corresponds to the recency scores of the given items in $x_u$. $\beta$ controls how much importance is given to the KL term, $z_u$ is a variational parameter of the variational distribution $\theta$ and $\phi$ are model parameters.

\subsection{Multi-Objective Optimization}
Optimizing a recommender on multiple objectives is non-trivial. Thanks to the recent work of \cite{milojkovic2019multi}, we employ the proposed multi-gradient descent algorithm for multiple objectives to train our recommenders. 
After a standard forward pass, the loss and gradient are computed for each objective and weights of the gradients are computing as a Quadratic Constrained Optimization Problem \cite{desideri2012multiple}. This can be solved analytically for two objectives, or solved as a constrained optimization problem as proposed in \cite{sener2018multi} for more than two objectives. Solving it allows us to obtain the common descent vector and update the parameters.
This training procedure enables us to incorporate both our temporal context and the relevance objectives to retrieve time-aware recommendations. The algorithm adapts the weight repartition between the two objectives in an advanced manner to optimize both during training.

\section{Experiments}
\label{ExperimentsSection}
\subsection{Datasets}

We study the performance of various models on three real-world publicly available datasets. 

\paragraph{MovieLens-20M.} Contains about 20 million ratings\footnote{\url{https://grouplens.org/datasets/movielens/20m/}.}, with values between 1 and 5. 
We binarize the user-item interaction matrix, keeping ratings of 4 and above as positive feedback to transform it into implicit feedback. We filter out all users with less than five ratings, and all movies rated by less than five users.
We focus on the last ten years of available data (2005-2015). The preprocessed dataset contains 46,295 users, 9479 items, and 3.76M interactions with a density of~0.86\%.

\paragraph{Steam.} Has review information from the gaming platform Steam\footnote{\url{https://cseweb.ucsd.edu/~jmcauley/datasets.html}.}. We converted user-item interactions into a positive feedback signals.
The dataset contains reviews from 2010 to 2018; however, the platform only sees an uptick in review activity after 2014, therefore we use 2014-2018 for our analysis. After preprocessing the dataset contains 471,457 users, 13,018 items, and 3.14M interactions with a density of 0.09\%.

\paragraph{Netflix.} The well-known Netflix Prize Competition dataset\footnote{\url{https://www.kaggle.com/netflix-inc/netflix-prize-data}.}. It consists of over 100 million ratings. 
We filter these ratings in the same way as the MovieLens ratings, and take the last two years of activity (2003-2005). Because of low performance on certain baselines, we denote two variants for the implicit feedback: one with threshold of~$4$ and above (Netflix$\geq$4), the other one with a threshold of~$5$ (Netflix$\geq$5). After preprocessing the dataset contains 257,775 users, 13,995 items, and 38.87M interactions with a density of 0.59\%.

\subsection{Recommendation Techniques}
We conduct experiment with the following well-known recommendation systems:

\paragraph{Mult-VAE\textsuperscript{PR}.} The MAMO framework\footnote{\url{https://github.com/swisscom/ai-research-mamo-framework}.} and the setup from the original paper \cite{liang2018variational} are utilized.

\paragraph{SVD. } The PyTorch implementation\footnote{\url{https://pytorch.org/docs/stable/generated/torch.svd.html}.} of the Singular Value Decomposition \cite{sarwar2000application} is used, taking the top 100 dimensions.

\paragraph{NCF. } The Neural Collaborative Filtering \cite{he2017neural}, we take the implementation from \footnote{\url{https://github.com/guoyang9/NCF}.}, sample 4 negative instances for every existing user-item interaction, set the predictive factor of 64, and the number of hidden layers for the multilayer perceptron (MLP) to three.
We do not present results obtained using pre-trained NeuMF, as they exhibited the same patterns as generalized matrix factorizaion (GMF) and MLP, but did not give a significant improvement.
To resolve difficulties in obtaining good results with \emph{Netflix$\geq$4} for GMF and MLP models, we used instead the \emph{Netflix$\geq$5} dataset.

\paragraph{BERT4Rec. } We implement this sequence-aware recommender system from \cite{sun2019bert4rec} in PyTorch and integrate it with the MAMO framework.
We take this model to show how directly encoding temporal information in the model impacts performance. In this case the ordering represents the temporal information.
Hyperparameters were mostly taken from the original paper, otherwise selected based on a simple grid search. 
The number of transformer layers is set to 2, the head number is 4, head dimensionality is 64, and the dropout is 0.1.
We use a sequence length of 100, while the proportion of masked inputs is 0.2.
The model is trained using the Adam optimizer with a learning rate of 1e-4.
All models were trained with the Adam optimizer, with a learning rate of 0.001.

\subsection{Experimental Setup}

We explore whether validation set formation in the  \emph{deployment-ready} phase may lead to false confidence in the performance of the evaluated model.
In the \emph{deployment-ready} phase, what we call the validation set is not necessarily used for hyperparameter tuning, but to assess the performance of the model before it is deployed.
There are minor differences in the datasets used for models with and without user representation. Models without user representation require some input interactions to be able to predict targets, while those with simply need to be passed a user identifier.
We divide our experiments into three sets, corresponding to the type of evaluation.

\subsubsection{Traditional Evaluation.}
Similarly to \cite{liang2018variational}, we divide the users 80:10:10 to form a train, validation, and test set.
The target interactions are selected by randomly sampling 20\% of the user-item interactions in the validation and test sets.

We show that if a model is evaluated on and then applied to a task that entails predicting randomly held-out interactions, the performance achieved on both validation and test sets is comparable. 
This traditional approach is typically used to report model performance.

We then contrast performance on randomly held-out interactions in the validation set against temporally held-out interactions in the test set. 
We take 5\% of the users from the train set to create the validation set and randomly hold-out 20\% of their interactions.
The train and validation sets contain user-item interactions up to a specific point in time.
The test set contains the interactions and users from the train and validation sets as inputs, and the temporally held-out interactions are targets, to simulate deployment in a commercial setting.

\subsubsection{Temporal Evaluation.}
We show that when evaluated with either a proportional or hard temporal cutoff, the model's performance is closer to what would be observed in a real-life setting.
However, it is important to note the ideal evaluation technique is heavily domain dependent.

We divide the train, validation, and test sets as follows. 5\% of the users from the train form the validation set.
In the first approach, we hold out the last 20\% of user-item interactions from each user in the validation set. While in the other, we hold out the last couple of months of activity and evaluate the model's ability to predict these interactions.
The test set contains the interactions and users from the two other sets as inputs, and the temporally held-out interactions are targets.

\subsubsection{Temporal Evaluation with Added Temporal Context.}
We introduce temporal context into the traditionally time-independent Mult-VAE\textsuperscript{PR} by using the work from \cite{milojkovic2019multi} to optimize the model for accuracy and recency. 
To calculate the recency score we take the timestamp of the moment that the item first became available, or the first recorded instance of any user interacting with the given item. This timestamp is $t_x$ in the recency function \ref{eq:recency}.
The strict temporal cutoff validation set is utilized, as well as the temporal test set described previously.

\subsection{Evaluation Metrics}
We evaluate models using three ranking metrics, as RS can often only show a predefined number of recommendations. We ensured that the items that the user had previously interacted with were removed from the output before the top-k results were selected for metric calculation. 

\begin{itemize}
\item \emph{Precision@K: }calculates how many of the recommended items are relevant to the user;

\item \emph{Recall@K: }quantifies the proportion of relevant items in the top-k recommended items by calculating how many of the desirable items are are suggested to the end-user. We take our definition from \cite{liang2018variational};

\item \emph{Recency@K: }assigns a recency score to each item, calculating the rating of the top-k recommended and relevant items. For user $u$ with relevant items $I_{u}$ we define $\omega(k)$ as the item at rank $k$, where $\mathbb{I}$ is the indicator function: \begin{equation}
Recency@K(u, \omega, f) = \sum_{k = 1}^{K}{\mathbb{I}
[\omega(k) \in I_{u}]\times f(\omega(k))}
\end{equation}
\end{itemize}

\section{Results}

\subsection{Traditional Evaluation.}
This experiment aims to show that the traditional way of evaluation recommender systems, shown in Table \ref{table:vaeTraditional}, is not a faithful representation of the environments in which they are actually deployed. The good performance achieved by evaluating in this way can provide a false sense of security.

\begin{table}
 \caption{Results of initial Mult-VAE\textsuperscript{PR} experiments, evaluated on a traditional evaluation protocol. We report Recall / Precision at $k=20$.}
 \centering
\begin{tabular}{@{}lcc@{}} 
\toprule
Dataset & Val\textsuperscript{trad} & Test\textsuperscript{trad} \\
\midrule
ML-20M &  0.31 / 0.17   & 0.31 / 0.17 \\ 
Steam  &  0.20 / 0.02  & 0.20 / 0.02\\ 
Netflix$\geq$4 & 0.35 / 0.19 & 0.35 / 0.19 \\ 
\bottomrule
\end{tabular}
\label{table:vaeTraditional}
\end{table}

Our claim is supported by the values highlighted by Table \ref{table:combinedResults}. Even though the validation sets are not identical to the ones before, the performance observed is very similar. However, it degrades on the time delayed test set, or to be more precise, when we simulate what would happen in a production setting. Drops in performance of -65.63\%, -35.00\%, and -71.43\% can be observed, on the Recall@20 values. 
We postulate that this discrepancy leads to significant dissonance between the results of certain recommenders as reported in literature, and those observed in their real-life application.

\begin{table}
 \caption{\label{table:combinedResults}Results of the Mult-VAE\textsuperscript{PR}, SVD, GMF, and MLP evaluated on a traditional, proportionally selected temporal, and strict cutoff validation set, as well as on a temporally held out test set. 
 Results of BERT4Rec evaluated on  a strict cutoff validation set and a time delayed test set.
 We report Recall / Precision at $k=20$.}
\centering
\begin{tabular}{@{}llcccc@{}} 
\toprule
Dataset & Model& Val\textsuperscript{trad} & Val\textsuperscript{prop} & Val\textsuperscript{cutoff}& Test\textsuperscript{temp} \\
\midrule
\multirow{5}{*}{ML-20M} & Mult-VAE\textsuperscript{PR} &  0.32 / 0.18   & 0.26 / 0.13 & 0.11 / 0.06 & 0.11 / 0.07\\ 
& SVD & 0.25 / 0.22   & 0.14 / 0.11 & 0.07 / 0.03 & 0.11 / 0.07\\ 
& GMF & 0.25 / 0.22   & 0.11/ 0.10 & 0.08 / 0.03 & 0.10 / 0.07\\ 
& MLP & 0.25 / 0.23   & 0.12 / 0.10 & 0.07 / 0.03 & 0.11 / 0.07\\ 
& BERT4Rec & -    & -  & 0.20 / 0.09   & 0.15 / 0.08 \\ 
\midrule
\multirow{3}{*}{Steam} & Mult-VAE\textsuperscript{PR} &  0.20 / 0.02   & 0.14 / 0.02 & 0.11 / 0.01 & 0.13 / 0.01\\ 
& SVD & 0.10 / 0.02   & 0.10 / 0.02 & 0.09 / 0.01 & 0.08 / 0.01\\ 
& BERT4Rec & -    & -  & 0.21 / 0.02  & 0.17 / 0.02\\ 
\midrule
\multirow{3}{*}{Netflix$\geq$4} & Mult-VAE\textsuperscript{PR} &  0.35 / 0.18   & 0.22 / 0.10 & 0.12 / 0.05 & 0.10 / 0.05\\ 
& SVD & 0.23 / 0.16   & 0.23 / 0.16 & 0.09 / 0.05 & 0.07 / 0.04\\
& BERT4Rec & -    & -  & 0.24 / 0.13 & 0.20 / 0.05\\ 
\midrule
\multirow{3}{*}{Netflix$\geq$5} & SVD & 0.23 / 0.10   & 0.23 / 0.11 & 0.12 / 0.05 & 0.09 / 0.03\\ 
& GMF & 0.31 / 0.14   & 0.30 / 0.14 & 0.14 / 0.05 & 0.12 / 0.04\\ 
& MLP & 0.31 / 0.14   & 0.30 / 0.14 & 0.14 / 0.05 & 0.12 / 0.04\\ 
\bottomrule
\end{tabular}
\end{table}

\subsection{Temporal Evaluation.}
The results shown in Table \ref{table:combinedResults} depict what happens when using traditional validation as opposed to our proposed evaluation sets. 
The table illustrates how the strict cutoff validation set approximates the \emph{deployment} behavior.
For all datasets, this approach seems to be a closer estimation of the "real-life" performance.
For example, the drop in performance is reduced from -71.43\% to -16.67\% on the Netflix$\geq$4 dataset for the Mult-VAE\textsuperscript{PR} model.
The proportionally selected validation sets seems to work well for the Steam dataset, and we know from industry experience that it can be good on others. However, this seems to be highly dataset specific.

Table \ref{table:combinedResults} also shows that this phenomenon is not isolated to the Mult-VAE\textsuperscript{PR}, but can be repeated with the SVD, GMF, and MLP models. As mentioned before, we were unable to conduct experiments on Netflix$\geq$4 with the GMF and MLP models; therefore we report their results on Netflix$\geq$5.
It is important to note that simpler methods, especially those based on matrix factorization, do not deal well with the Steam dataset.
This is the sparsest dataset that we work with
which seems to make it difficult to learn anything meaningful. 
Based on this, we exclude the Steam dataset for GMF and MLP.
However, we keep the results for SVD.

We strongly recommend that these evaluation methods be taken into account when presenting novel achievements in the field.
When feasible, we recommend to apply both protocols.

\subsection{Temporal Evaluation and Temporal Models.} The results presented so far were achieved using traditional recommender architectures, with no way of learning temporal dynamics. To show that it is possible to achieve better results on the given datasets, we incorporate the temporal dynamics into the training process, by utilizing BERT4Rec. The results are shown in  \ref{table:combinedResults}, and dominate all traditional solutions.
This confirms our hypothesis that temporal dynamics should be accounted for in both evaluation design and model architecture in order to attain the best possible recommenders.

With BERT4Rec the interaction ordering is encoded in the model. The authors acknowledge that the naive recency objective does not reproduce this effect when added to traditional RS. However, the goal of the subsequent subsection is to illustrate that the simple addition of a cheap time-dependant weight affects performance in a meaningful way.

\subsection{Temporal Evaluation with Added Temporal Context.} To integrate the temporal context into the traditional models, our following contribution has the recency included as an objective influencing the optimization.
We refer to the multi-objective Mult-VAE\textsuperscript{PR} as the Multi-Objective Recency Enriched mult-VAE\textsuperscript{PR}(MOREVAE).

\begin{table}
\caption{\label{table:vaeRecncyAdded}Comparison of Mult-VAE\textsuperscript{PR} and MOREVAE results obtained on temporally held out test sets. We report \textbf{R}ecall, \textbf{P}recision, and \textbf{Re}cency at $k=20$.}
\centering
\begin{tabular}{@{}llccc@{}} 
\toprule
Dataset & Model & R & P & Re\\
\midrule
\multirow{2}{*}{ML-20M} &  Mult-VAE\textsuperscript{PR} & 0.11   & 0.07 &0.23 \\ 
& MOREVAE     & \textbf{0.13}   & \textbf{0.08} & \textbf{0.47}\\
\midrule
\multirow{2}{*}{Steam}  & Mult-VAE\textsuperscript{PR} & \textbf{0.13}  & \textbf{0.01}& 0.15\\ 
& MOREVAE     & \textbf{0.13}   & \textbf{0.01} & \textbf{0.18}\\
\midrule
\multirow{2}{*}{Netflix$\geq$4} & Mult-VAE\textsuperscript{PR} & 0.10 & 0.04 &0.34\\ 
& MOREVAE     & \textbf{0.12}  &  \textbf{0.05} & \textbf{0.66} \\ 
\bottomrule
\end{tabular}
\end{table}

\begin{figure*}[t]
\centering
\subfloat[ML20m dataset.\label{fig:sub1}]{ \includegraphics[width=.3\linewidth]{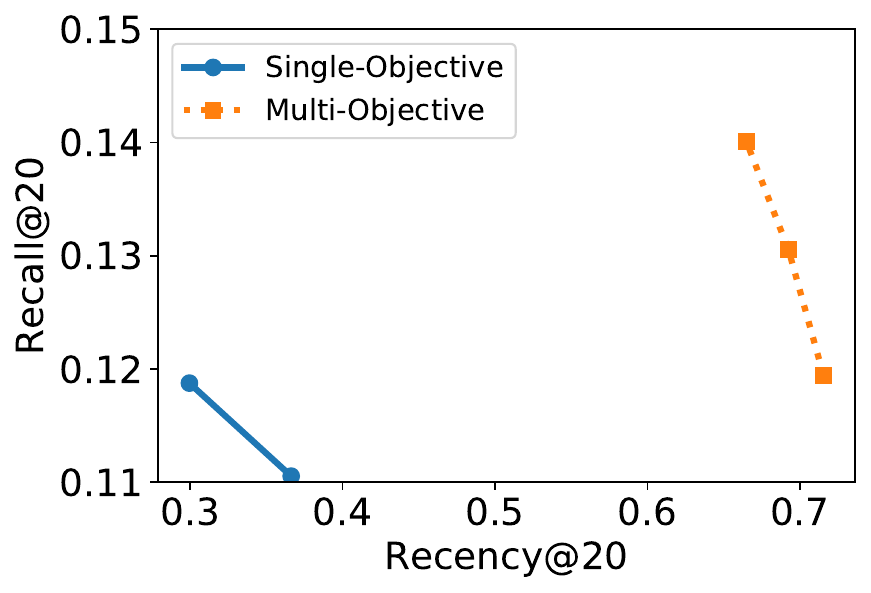}}
\hfill
\subfloat[Steam dataset.\label{fig:sub2}]{ \includegraphics[width=.3\linewidth]{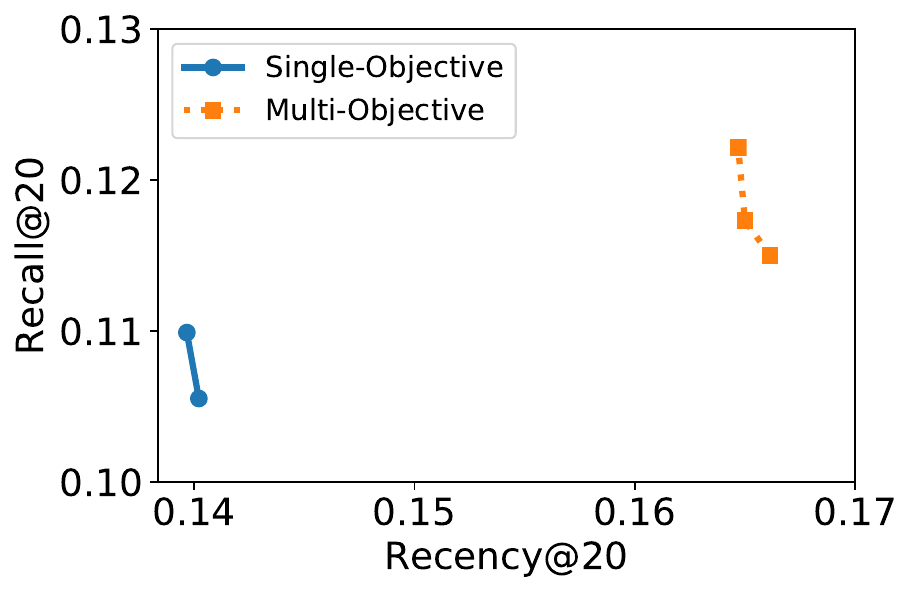}}
\hfill
\subfloat[Netflix$\geq$4 dataset.\label{fig:sub3}]{ \includegraphics[width=.3\linewidth]{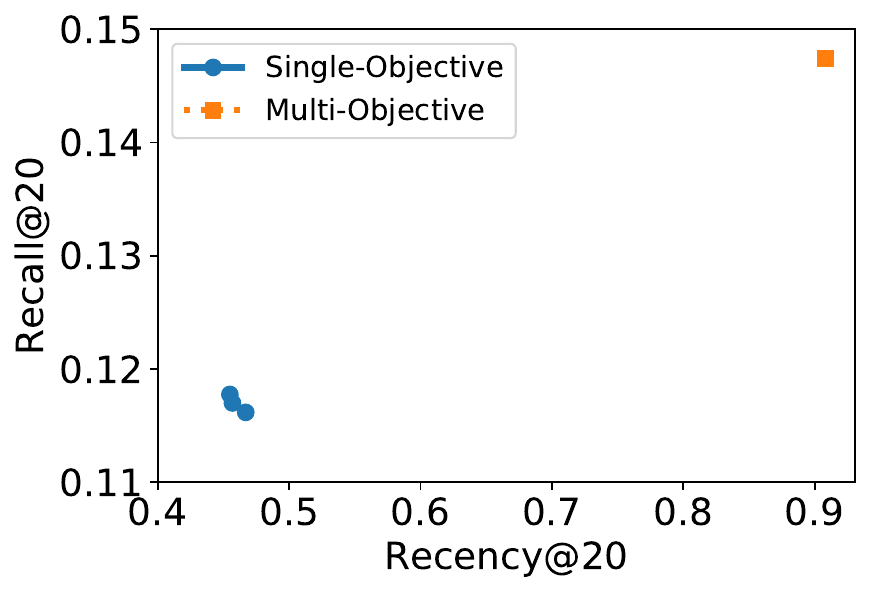}}

\caption[Comparison of Pareto Fronts obtained through optimizing on one objective and two objectives]{Pareto Fronts obtained through optimizing on one objective (accuracy), and two objectives (accuracy and recency).}

\label{fig:paretofronts}
\end{figure*}

We present both the Pareto Fronts obtained during training and the results of the best models on the test sets. These results were obtained through more intense training than those shown in the previous sections in an attempt to extract the best possible performance from the Mult-VAE\textsuperscript{PR}.
The Pareto Fronts were generated by evaluating on the strict cutoff validation sets during training, and the best models were chosen by selecting those with the highest Recall@20 and applying them to the time delayed test sets. 
Figure \ref{fig:paretofronts} shows that the multi-objective approach not only dominates the single objective one in terms of recency, but that optimizing for recency also increases the relevance of the recommendations, validating our initial intuition.
The results of the best models over the test sets are shown in Table \ref{table:vaeRecncyAdded}. 
The improvements obtained are 18.18\%, 0.00\%, and 20\% for Recall@20; 14.29\%, 0.00\%, and 25.00\% for Precision@20.
The improvements seen in Recency@20 are 104.35\%, 20.00\%, and 94.12\%.

\section{Conclusion}



Following standard offline recommendation evaluations during development 
may lead to false confidence when deploying models in real-life scenarios. 
Utilizing random sampling to hold out data is not an adequate approximation of many real-life use-cases.
Previous research generally focuses on developing better metrics to reflect real-world performance, but still omits temporal context. We highlight this lack of standardization and propose two temporal evaluation protocols that empirically better approximate real-life conditions.

Our second contribution is to propose leveraging a multi-objective approach and train models on relevance and recency simultaneously. We show that a naive recency objective can be used to integrate temporal information in existing time-unaware recommenders. 
Experiments on three real-world publicly available datasets demonstrate that our method produced solutions that strictly dominate those obtained with a model trained on a single-objective optimization.

We explored datasets that are frequently used in recommender systems research, all related to digital media content. Digital media content is consumed frequently and generally without much repetition. The importance of recency and capturing transient behavioral trends may not be equivalent in other recommender systems applications, such as grocery or clothes shopping. The influence of temporal dynamics on these sectors is an exciting topic, and we leave it to future academic and commercial research.

\end{document}